\documentclass[eqsecnum,aps,ams,preprint,epsf]{revtex4}
\usepackage{graphicx}
\usepackage{wrapfig}
\font\tenbifull=cmmib10 scaled 1200 
\font\tenbimed=cmmib9
\font\tenbismall=cmmib7
\def\bmit{\fam9 }
\mathchardef\bbgamma="710D
\mathchardef\bbkappa="7114
\mathchardef\bbrho="711A
\mathchardef\bbsigma="711B
\mathchardef\bbtau="711C
\mathchardef\bbvarrho="7125
\mathchardef\bbvarsigma="7126
\mathchardef\bbPhi="7008
\mathchardef\bbxi="7118
\def\boldgamma{{\bmit\bbgamma}}

\def\boldtau{{\bmit\bbtau}}

\def\boldPhi{{\bmit\bbPhi}}

\def\boldeta{{\mbox{\boldmath$\eta$}}}

\newcommand{\be}{\begin{eqnarray}&&}
\newcommand{\ee}{\end{eqnarray}}

\def\dfrac{\displaystyle\frac}
\def\la {\langle\,}
\def\ra {\rangle\,}
\begin{document}

\newpage

\thispagestyle{empty}
\title{$\bf \eta'$ meson production in nucleon-nucleon collisions\\
near the threshold}

\author{L.P. Kaptari}
\altaffiliation{On leave of absence from
Bogoliubov Lab. Theor. Phys. 141980, JINR,  Dubna, Russia}
\author{ B. K\"ampfer}
\affiliation{Forschungszentrum Dresden-Rossendorf, PF 510119, 01314 Dresden, Germany}

\date{\today}

\begin{abstract}
The production of $\eta'$ mesons in the
reactions $pp\to pp\eta'$ and $pn\to pn\eta' $ at threshold-near energies
is analyzed within a covariant effective meson-nucleon theory.
The description of
cross section and angular distributions
of the available data in this kinematical region
in the $pp$ channel is accomplished
by including meson currents and nucleon currents  with the
resonances $S_{11}(1650)$, $P_{11}(1710)$ and $P_{13}(1720)$.
Predictions for the $pn$ channel are given.
The  di-electron production  from subsequent $\eta'$ Dalitz decay
$\eta' \to \gamma \gamma^* \to\gamma e^+e^-$
is also calculated and
numerical results are presented for intermediate energy
and kinematics of possible experiments with HADES, CLAS and KEK-PS.
\end{abstract}
\maketitle

\section{Introduction}

The pseudo-scalar mesons $\eta$ and $\eta'$
represent a  subject of considerable interest since some time
which has been addressed  in many investigations
(see, e.g.\ \cite{diekmann,feldmann_PSmesons,christos_U(1),manual_U(1),LandPhysRep}).
The physical states of   $\eta$, $\eta'$ mesons can be constructed as
a superposition of $\eta_0$, $\eta_8$ members of a SU(3)
pseudoscalar nonet with one mixing angle.
If the coupling constants (defined usually by the
relation $\la 0|J^\alpha_{\mu 5}|  P(p)\ra=if^\alpha_p p_\mu$,
where $P$ is a member of the SU(3) nonet, and $\alpha=0-8$)
follow the same pattern as the state mixing then they must be
related to each other as \cite{eta_etapr_mixing}
$
f^8_\eta=f_8\cos\theta, \,  f^0_\eta=-f_0\sin\theta, \,  
f^8_{\eta'}=f_8\sin\theta, \,  f^0_{\eta'}=f_0\cos\theta,
$
where $\theta$ is the state mixing angle.
A  combined phenomenological
analysis \cite{mixingangle} of data on two-photon decays
of  $\eta$ and  $\eta'$ as well as
$\gamma \eta$ transition form factors
shows  indeed  that one can parameterize the relation among the coupling constants
by such a relation, however, with an angle essentially differing from the state mixing angle.
The interrelation between the
two angles and different decay constants
can be found by considering the divergences of axial vector currents
including the $U(1)$ axial anomaly, which couples
the $\eta'$ state
with gluon fields \cite{witten_veneziano,japanAnomaly}.

The $U(1)$ axial anomaly  plays also a crucial role in understanding the physical masses
of the $\eta$ and $\eta'$ mesons from the
point of view of spontaneous breaking of chiral symmetry.
It is known \cite{christos_U(1),manual_U(1)} that the
non-zero divergence of the
axial current   $J_{\mu\,5}^0$ in the chiral limit 
in presence of interactions
is due to couplings with gluons arising after renormalization.
This anomaly is responsible for the generation of the heavy mass of the
physical $\eta'$ meson. Yet, the anomaly is connected with a violation of the
OZI rule for the nucleon and seems also to point to a small strangeness
admixture in the proton wave function \cite{etaprime_OZI}.

This also has a direct consequence in deep-inelastic
scattering. For instance, the Goldberger-Treiman relation (see, e.g.,
\cite{etaprime_OZI} and references therein) relates
the nucleon-nucleon-$\eta'$ coupling constant $g_{NN\eta'}$
(which is non-zero because of OZI rule violation)
 with the flavor-singlet axial constant $G_A(0)$ and to the coupling $g_{NNg}$
with gluons. If the gluon contribution is ignored in this relation then the EMC measurements
lead to the so-called "spin crisis" \cite{spincrisis}. A knowledge of all terms in
the Goldberger-Treiman would allow to understand the origin
of the spin crisis. However, neither $g_{NN\eta'}$ nor $g_{NNg}$ have been measured directly
in experiments. From this point of view an investigation  of the  $pp\to pp\eta'$ reactions
can be considered  a tool for supplying additional information on a wide
range of fundamental theoretical issues.

The strong decay channels of $\eta$ and $\eta'$ are highly suppressed
by various constraints (such as $G$ parity) so that their
widths are extremely narrow   implying a relatively long life-time
and, consequently, a suitable means  to investigate different theoretical aspects of
the  OZI rule, $U(1)$ axial anomaly, role of gluons in the $NN\eta'$ vertex etc.
A further remarkable fact is that near the threshold
the invariant mass of the $NN\eta'$ system
in corresponding  reactions is in the region of heavy nucleon resonances, i.e.\
resonances with isospin $\frac12$ can be investigated via these processes.
Furthermore, resonances weakly bound with nucleons (the so-called "missing resonances"
\cite{missingRes,manley}) can be studied in principle. First SAPHIR data on $\eta'$
photo-production \cite{SAPHIR_res} has shown a strong rise of the cross section
near the threshold which indicates a possibly large contribution from nucleon
resonance excitation. Moreover, if one assumes that the production mechanism is solely
governed by resonances
then the new states $S_{11}(1897)$ and $P_{11}(1986)$ can be predicted \cite{SAPHIR_res}.
However, further analysis of the SAPHIR and CLAS \cite{CLASnew_getaprimenn,CLASold}
data has shown \cite{nakaym_gnneta,zhang,lothar} that data are also
compatible with a larger number of
resonances, including  higher spin states $\frac32$ and $\frac52$, i.e., data are still
too scarce to firmly determine the properties  of the relevant resonances.

Another aspect of $\eta'$ production in elementary hadron reactions is
that the subsequent Dalitz decay may constitute a prominent source
of di-electrons in intermediate-energy heavy-ion collisions.
Indeed, the recent HADES data \cite{hades07} exhibit a sizeable di-electron
yield at large invariant masses
which, besides the contribution from
the vector mesons $\rho$ and $\omega$ \cite{barz},
can be due to $\eta/\eta'$ mesons too. The $\eta$ Dalitz decay  has been
quantified \cite{barz} as prominent contribution to the invariant mass spectrum, indeed.
One of the primary aims of the HADES experiments \cite{HADES} is to seek
for signal of chiral symmetry restoration in compressed nuclear matter. For such an
endeavor one needs a good control of the background processes, including
the $\eta'$ Dalitz decay, in particular at higher beam energies,
as becoming accessible at SIS100 within the FAIR project \cite{FAIR}.

The $\eta'$ Dalitz decays depend on the pseudo-scalar transition form factor, which
encodes hadronic information   accessible in first-principle QCD calculations or
QCD sum rules~\cite{LandPhysRep}.
The Dalitz decay process of a pseudo-scalar meson $P$ can be presented as
$
P\to \gamma +\gamma^* \to \gamma  + e^- + e^+.
$
Obviously, the probability of emitting a virtual photon is
governed by the dynamical electromagnetic
structure of the "dressed" transition vertex $P\to \gamma^*$ which
is encoded in the  transition form factors
\cite{LandPhysRep,feldmann_FF,CLEO_FF,ourOmegaFF,ourEta}.
If the decaying particle
were point like, then calculations of $\gamma^*$ mass distributions
and decay widths  would be straightforward along standard quantum
electrodynamics (QED) techniques. Deviations of the measured quantities from
the QED predictions directly reflect the
effects of  the form factors  and, thus the internal hadron structure,
and, consequently, can serve as experimental tests  to
discriminate  the different theoretical  models in the non-perturbative QCD 
regime. Yet, calculations of the
transition form factor within perturbative QCD involve triangle and box diagrams,
which are tightly connected to the  $U(1)$ axial anomaly so that form factors provide additional
information on the nature of anomalies in quantum field theories, here QCD.

In the present paper we study the di-electron production from
Dalitz decay of   $\eta'$ mesons in $pp$ and $pn$ reactions
at beam energies of a
few GeV for kinematical conditions corresponding to the
HADES setup \cite{HADES}. Our focus is to elaborate a model which provides a reliable energy
dependence of the total cross section of $\eta'$ meson production and
its subsequent decay into a di-electron and a photon at threshold-near  beam energies.
To this end we calculate the dependence of the differential cross section
for the reaction $pp\to pp\gamma e^+e^-$
upon  the invariant mass of the subsystem $\gamma e^+e^-$ around the pole masses
of $\eta'$.
We calculate the invariant mass distribution of di-electrons
in a suitable kinematical range
as a function of the di-electron invariant mass and argue
that such a quantity, normalized to the real photon point and
supplemented by some specific kinematical factor,
represents the transition form factor. In such way the extraction of the
transition form factor becomes accessible.

$\eta'$ production processes have been analyzed in several papers.
In Refs. \cite{nakaym_gnneta,zhang,li,lothar},
photo-production of $\eta '$ has been studied to elucidate
the role of resonances.  An investigation of the $\eta'$ production in $NN$ reactions
has been performed in Refs.~\cite{nakayama_eta,nakayam_etapr}, and a consistent combined
analysis of $\eta'$ production in the reactions  $pp\to pp\eta'$ and
$\gamma p\to  p\eta'$  has been attempted in Ref.~\cite{nakayam_consist}.

Our paper is organized as follows.
Section II is devoted to the theoretical background for
dealing with the reaction $pp \to pp \eta' \to pp \gamma e^+ e^-$.
It is essentially based on the effective
model \cite{ourOmega,ourOmegaFF,ourEta,ourNuclPhys}
proposed to describe bremsstrahlung and
vector meson production in $NN$ reactions. The model
is based on direct calculations of the relevant
Feynman diagrams within a covariant phenomenological meson-nucleon theory.
The  model parameters  have been  fixed from independent
experiments and adjusted to achieve a good description of the
available experimental data. In the present paper,
diagrams with excitation of  nucleon resonances with masses close
to the masses of a nucleon plus $\eta'$ meson have been also included.
These are $S_{11}(1650)$, $P_{11}(1710)$ and $P_{13}(1720)$ resonances.
The corresponding effective constants, whenever possible, have been
obtained from the known decay widths of  direct decay into the $\eta'$ channel or
radiative decay with subsequent use of vector meson dominance conjecture.
Also we use effective constants commonly adopted in the literature and
obtained from different considerations, e.g., SU(3) symmetry, fit of photo-absorbtion
reaction etc.\ \cite{nakayama_eta}. In Section III, total cross sections
and angular distributions for $\eta'$ production in
$NN$ reactions are presented. Comparison with available experimental data is also performed.
In Section IV, the $\eta'$ Dalitz decay is considered.
Results of calculations of the invariant mass distribution as a function of the di-electron
mass are presented for different initial  energies.
The role of the transition form factor in such processes is investigated as well.
Our conclusions are summarized in Section V.

\section{The model \label{subsecOdin}}

We consider the reaction
\begin{equation}
N_1 + N_2 \to N_1 + N_2 + \eta' \to N_1 + N_2 + \gamma +e^+ +e^-.
\label{dalitzpp}
\end{equation}
The invariant cross section is
\begin{eqnarray}
d^{11}\sigma =
\frac{1}{2\sqrt{\lambda(s,m^2,m^2)}}\frac{1}{(2\pi)^{11}}
\frac14 \sum\limits_{\rm spins}\
\,|\ T(P_1',P_2',k_1,k_2,k_\gamma,{\rm spins}) \ |^2 d^{11}\tau_f \ \frac{1}{n! },
\label{crossnn}
\end{eqnarray}
where
the factor $1/n!$ accounts for $n$ identical
particles in the final state, $\vert T \vert^2$ denotes the invariant amplitude
squared, $m$ is the nucleon mass, $s$ denotes the invariant mass
of a particle or a sub-system of particles  with total four-momentum  $p$, i.e., $s=p^2$,
and $d\tau_f$ is the invariant phase volume.
The kinematical factor $\lambda$,
describing the incident flux,
is defined as $\lambda(x^2,y^2,z^2)\equiv (x^2-(y+z)^2)(x^2-(y-z)^2)$.
Eventually, $P_1',P_2',k_1,k_2,k_\gamma$ are
the momenta of the final nucleons, leptons and photon, respectively.

\subsection{Interaction Lagrangians}

The invariant amplitude $T$ is evaluated here
within a phenomenological meson-nucleon  theory based on
effective interaction Lagrangians
(see \cite{nakayama_eta,ourOmega,ourOmegaFF,ourEta,ourNuclPhys})
which include
(i) isoscalar mesons: scalar ($\sigma$), pseudo-scalar ($\eta$), vector ($\omega$),
(ii) isovector mesons: scalar  ($a_0$), pseudo-scalar  ($\pi$) and vector ($\rho$):
\begin{eqnarray}
{\cal L}_{\sigma NN }&=& g_{\sigma NN} \bar N  N \it\Phi_\sigma , \label{mnn1}\\
{\cal L}_{a_0 NN }&=& g_{a_0 NN} \bar N (\boldtau \boldPhi_{a_0}) N \it , \label{mnn2}\\
{\cal L}_{\pi NN}&=&
-\frac{f_{ \pi NN}}{m_\pi}\bar N\gamma_5\gamma^\mu \partial_\mu
({\boldtau \boldPhi_\pi})N , \label{mnn4}\\
{\cal L}_{\eta NN}&=&
-\frac{f_{\eta NN}}{m_\eta}\bar N\gamma_5\gamma^\mu \partial_\mu
\Phi_\eta N , \label{mnn5}\\
{\cal L}_{\rho NN}&=&
-g_{\rho NN }\left(\bar N \gamma_\mu{\boldtau}N{\boldPhi_ \rho}^\mu-\frac{\kappa_\rho}{2m}
\bar N\sigma_{\mu\nu}{\boldtau}N\partial^\nu{\boldPhi_\rho}^\mu\right) ,\label{mnn3}\\
{\cal L}_{\omega NN}&=&
-g_{\omega  NN }\left(
\bar N \gamma_\mu N {\it\Phi}_{\omega}^\mu-
\frac{\kappa_{\omega}}{2m}
\bar N \sigma_{\mu\nu}  N \partial^\nu \it\Phi_{\omega}^\mu\right),
\label{mnn}
\end{eqnarray}
where $N$ and $\it\Phi$ denote the nucleon and meson fields, respectively,
and bold face letters stand for iso-vectors.
All couplings with off-mass shell
particles are dressed by monopole form factors
$F_M=\left(\Lambda^2_M-\mu_M^2\right)/\left(\Lambda^2_M-k^2_M\right)$,
where $k^2_M$ is the 4-momentum of a virtual particle with mass $\mu_M$.

To describe the Dalitz decay of the $\eta'$ meson Eqs.~(\ref{mnn1})-(\ref{mnn})
must be supplemented with the corresponding Lagrangians
describing the interaction of the electromagnetic field $A_\mu$ with
di-electrons ($ll = e^+e^-$) and with the $\eta'$ meson,
\begin{eqnarray}
{\cal L}_{\gamma ll }&=&-e(\bar\psi_l \gamma_\mu\psi_l)A^\mu,\label{mnn8}\\
{\cal L}_{\eta'\gamma\gamma }&=&
f_{\eta'\gamma\gamma}\left( \varepsilon_{\alpha\beta\mu\nu}
\partial^\beta A^\alpha \partial^\nu A^\mu
\right) \Phi_{\eta'},
\label{em}
\end{eqnarray}
where $\varepsilon_{0123}=-1$. Lagrangians (\ref{mnn1})-(\ref{em})
determine the $S$ matrix from which one  generates
the corresponding tree-level Feynman diagrams
describing the "bremsstrahlung" of $\eta'$ off  nucleons
and the subsequent Dalitz decay $\eta'\to \gamma e^+e^-$. Usually such diagrams are
called in the literature as nucleonic currents in Dalitz decay of mesons.

\begin{figure}[h]  
\includegraphics[width=1.\textwidth]{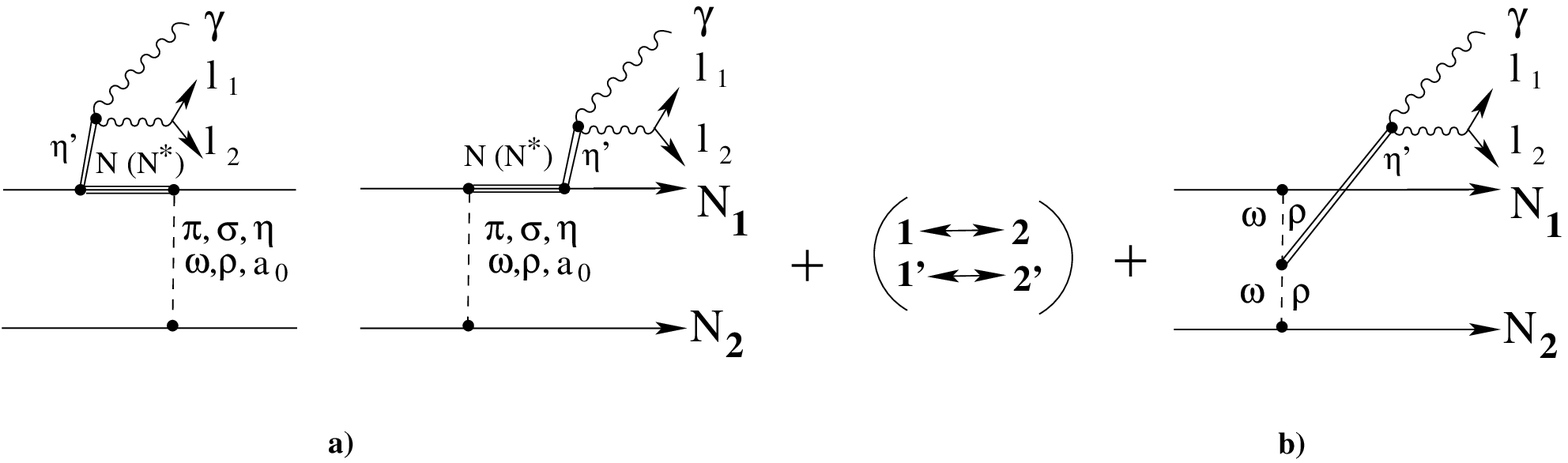}
\caption{Tree-level
diagrams for the process $NN\to NN\gamma e^+e^-$ within an effective
meson-nucleon theory. a) Dalitz Decay of
$\eta'$-mesons from nucleonic and resonance currents.
b) Dalitz decay of $\eta'$ mesons from  internal meson conversion.}
\label{fig0}
\end{figure}

The  $\eta'$ meson can  be produced also by an internal conversion
of the exchanged mesons, the so-called conversion or meson current.
The dominant exchange mesons in this case
are $\omega$ and
$\rho$ mesons with the interaction Lagrangians
\begin{eqnarray}
{\cal L}_{\eta' \omega\omega}&=& -\frac{g_{\eta \omega\omega}}{2m_\omega}\,
\varepsilon_{\mu\nu\alpha\beta}
\,\left (\partial^\mu\it \Phi_\omega^\nu    \partial^\alpha\it
\Phi_\omega^\beta \right)\Phi_{\eta'},
\label{conversion1}\\
{\cal L}_{\eta' \rho\rho}&=& -\frac{g_{\eta' \rho\rho}}{2m_ \rho}\,
\varepsilon_{\mu\nu\alpha\beta}
\,\left (\partial^\mu\it \boldPhi_ \rho^\nu    \partial^\alpha\it
\boldPhi_ \rho^\beta \right)\Phi{_\eta'}.
\label{conversion}
\end{eqnarray}
The corresponding diagrams are the same as in $\eta$ production \cite{ourEta}, see
Fig.~\ref{fig0}.

In the threshold-near kinematics for the $\eta'$ production in $NN$
reactions there are several firmly determined nucleon resonances
(four stars, according to the Particle Data Group classification \cite{pdg})
and a  large number of the
so-called "missing  resonances" \cite{missingRes,manley},
which, in principle, can be excited in these processes.
Consequently,  $\eta'$ production can serve as
additional tool to search for and to investigate  the  missing resonances
in the mass interval $1.5 - 2.0\,  GeV/c^2$. As mentioned in the Introduction,
in Refs.~\cite{nakayam_etapr,zhang,li}
a preliminary analysis of SAPHIR data \cite{SAPHIR_res} has been performed
assuming that only resonance currents contribute in this kinematical region. Then,
excitations of  missing  $S_{11}$ and $P_{11}$ resonances have been considered
and their pole masses and widths are estimated.
A further combined analysis~\cite{lothar},  including also
the CLAS data \cite{CLASnew_getaprimenn,CLASold},
has demonstrated that excitations of only two resonances
are not sufficient to describe the data,
and  at least two more missing  resonances,
$P_{13}$ and $D_{13}$, ought to be involved into the calculations. 
Eventually, a systematic analysis
of the $\eta'$ production in photo and $NN$ reactions \cite{nakayam_consist}
has shown that an equally well description of data could be achieved  with several
 diverse  sets of diagrams, which  include different numbers of known and missing resonances.
This means  that up to now the available data is too scarce to determine
unambiguously the kind and characteristics of resonances contributing in
this kinematical region. Moreover, since  in photo
and $NN$ reactions one can observe excitations of different  resonances,
it is not mandatory  to analyze simultaneously photo and $NN$ data within the same set of
diagrams. We are interested in finding  reliable parameterizations of the
energy dependence of the total cross section and angular distributions near
the threshold. Consequently, in order to reduce the number of free parameters and
to avoid additional ambiguities, in the present analysis we consider only the
known (four stars)
lowest spin resonances $S_{11}(1650)$ with odd parity, and
$P_{11}(1710)$ and $P_{13}(1720)$ with even parity.
The corresponding nucleon-meson-resonance interaction Lagrangians
can be found in  Ref.~\cite{nakayama_eta}. It should be mentioned
that the inclusion of higher spin resonances
leads to additional uncertainties.
It is known that the Lagrangian for particles with spins $ s \geq \frac32$
possesses an additional symmetry, i.e.,
additional free parameters (see Ref.~\cite{davitsonDelta} and
references therein quoted). Thus, the invariance of the Lagrangian with respect to the point
transformation causes  an additional transformation parameter
$A$ and the of-mass shell parameter $z$.
In the present calculation we adopt $A=-1$ and  $z=-1/2$. Also, the choice of the form of
higher-spin  propagators has been a subject of discussion in the
literature \cite{ourNuclPhys,prop4,prop5,prop6,prop7} with respect to the choice
of the spin projector operator $P_{\frac32}$ (off mass shell or on
mass shell) and the order to write the product
of the energy projection
operator $\hat P_{N^*}+m_{N^*}$ with the spin projection operator $P_{\frac32}$
(only for on-mass shell particles these two operators commute). In the present paper we
take the spin-$\frac32$ propagator in the form \cite{ourNuclPhys}
\begin{eqnarray}
&&
\label{prop32}
S_{\frac32}^{\mu\nu}(P)=-i\frac{\hat P_{N^*}+m_{N^*}}{P^2-m_{N^*}^2}
P_{\frac32}^{\mu\nu} (P),
\end{eqnarray}
where the spin projection operator is of the  commonly adopted form in the 
Rarita-Schwinger formalism \cite{fron}
\begin{eqnarray}
&&
\label{spr32}
P_{\frac32}^{\mu\nu} (P)=  g^{\mu\nu}-\frac13\gamma^\mu\gamma^\nu
-\frac{2P^\mu P^\nu}{3m_{N^*}^2}
-\frac{1}{3m_{N^*}}\left( \gamma^\mu P^\nu-\gamma^\nu P^\mu\right).
\end{eqnarray}

\subsection{Fixing  effective parameters}\label{subsecII}

The coupling constants, masses and cut-offs for the dipole form factors
for the nucleon currents are taken the same as in the Bonn potential C \cite{bonncd}, 
except for the $\omega$  meson where the coupling   
is chosen as $g_{\omega NN}=11$ (see also comments in 
Refs.~\cite{nakayama_eta,nakayam_etapr}).
The choice of the magnitude of the coupling constant $g_{\eta' NN}$ is
still  matter of debate. The OZI rule would imply a small value of
$g_{\eta' NN}$, while the above mentioned $\eta$-$\eta'$ mixing conjecture
can relate the corresponding
coupling constants and express $g_{\eta' NN}$ via $g_{\eta NN}$ providing
in such a way  a relatively large $g_{\eta' NN}$ \cite{zhang,nakayam_etapr}.
An analysis based on
$SU(3)$ symmetry and $\eta$-$\eta'$ mixing angle, as
performed in  Ref.~\cite{nakayam_etapr}, has established
an upper limit of  $g_{\eta' NN}=6.1$. Note that in the analysis \cite{nakayam_etapr}
some 50\% of the pseudoscalar-pseudovector admixture in the
$NN\eta'$ Lagrangian has been adopted.
Subsequent investigations have shown that such a constant is too large, and a new upper limit
$g_{\eta' NN}< 2$ has been proposed \cite{nakaym_gnneta} for the pseudo-vector coupling.
Moreover, even values consistent with zero
are also admitted in attempts to estimate
the couplings and masses of possible excitations of the missing
resonances \cite{nakayam_consist,zhang}.
Evidently, the choice of a small value $g_{\eta' NN}$ implies a negligible contribution
of the nucleonic currents, therefore increasing the role of meson conversion
and nucleon-resonance currents in the $\eta'$ production.

In our calculations we take
$g_{\eta' NN}\simeq 1.42$ as reported in Ref.~\cite{feldmann_PSmesons} and
recently  confirmed by the CLAS data \cite{CLASnew_getaprimenn}.
The effective coupling constants for the resonance currents, whenever possible, have been
estimated from the known decay widths of  direct decay into $\eta'$ channels or
radiative decay with subsequent use of the vector meson dominance (VMD) conjecture.
The few remaining unknown cut-off parameters are taken either as constant
or are adjusted to the experimental data. 
These parameters are listed in  Tab.~\ref{tb1}.

\begin{table}[!ht]\caption{Resonance Parameters. For the spin-$\frac32$ resonance
$P_{13}$ the off-shell parameter is taken as $z=-1/2$ and the second coupling constant
 with vector mesons~\cite{nakayama_eta} is given in parenthesis.}\vskip 2mm
 \begin{tabular}{c  ccccc| ccccc| ccccc }\hline\hline
  \multicolumn{1}{c}{\phantom{vert}}
  &\multicolumn{5}{c| }{$S_{11}(1650)$}
  & \multicolumn{5}{ c| }{$P_{11}(1710)$}
  &\multicolumn{5}{c }{$P_{13}(1720)$} \\
  &       & $ g_{MNN^*}$  &&
                       $\Lambda [GeV]$  &
                              &$\,\,\, g_{MNN^*}$
                                       & &$\Lambda [GeV]$
                                               & && $\, \, g_{MNN^*}$ &
                                                         $\Lambda [GeV]$ \\\hline
  $\pi$   & &  1.47&  &1.2 & &  1.47 & &1.2&   &   &  0.2   & 1.2  & &  &  \\
  $\eta$  & &  0.7 &  &1.2 & &  1.9  & &1.2&   &   &  0.6   & 1.2  & &  &      \\
  $\omega$& &  2.8 &  &1.2 & &  6.28 & & 1.2&  &   &  10(2) &   1.2& &  &           \\
  $\rho$  & &  1.1 &  &1.2 & &  2.1  & &1.2&   &   & -10(-6)& 1.2  & &  &        \\
  $\eta'$ & & 1.18 &  &0.9 & &  1.4  & &1.2&   &   & 1.0    &  1.2 & &  &             \\
  \hline
\end{tabular}
\label{tb1}
\end{table}

All vertices with  off-mass shell nucleons and nucleon resonances are dressed with
(i.e. are to be multiplied by)  a form factor
\begin{equation}
F(p^2,\Lambda) = \frac{\Lambda^4}{\Lambda^4+(p^2-m^2)^2}.
\label{vershff}
\end{equation}

The coupling constants  $g_{\omega\omega\eta'}$  and
$g_{\rho\rho\eta'}$ for the meson conversion current have been derived
from a combined analysis of
radiative decays within the vector meson dominance (VMD) model and within effective
Lagrangians with SU(3) symmetry providing
\begin{equation}
\label{omom}
g_{\omega\omega\eta'} \simeq 4.90, \quad g_{\rho\rho\eta'}\simeq 5.84.
\end{equation}

Note  that naive direct calculations of these constants within the VMD  model alone
can provide slightly larger values, e.g.  $g_{VV\eta'}\sim 6.5-7.0$. The corresponding
cut-offs,  $\Lambda_{\omega\omega\eta'}=\Lambda_{\rho\rho\eta'}=1.2\, GeV$, determine
the form factor in the conversion vertex
$\eta' VV$  (where  $V$ denotes $\omega$ or $\rho$)
which is chosen as (see  Refs.~\cite{nakayam_etapr,ourEta})
\begin{equation}
\label{convFF}
F_{VV\eta'}(\Lambda,k_1^2,k_2^2)=
\frac{\Lambda^2-m_V^2}{\Lambda^2-k_1^2}\frac{\Lambda^2}{\Lambda^2-k_2^2}.
\label{ffconv}
\end{equation}
In accordance  with the procedure of determining  the
coupling constant,
the form factor (\ref{ffconv}) is normalized to unity  if one vector meson is on-mass shell,
while the other one
becomes massless, e.g., $F_{VV\eta}(\Lambda,k_1^2=m_V^2,k_2^2=0)=1$.  
In calculations performed in
Ref.~\cite{nakayam_consist} the form factor (\ref{ffconv}) has been changed
from the monopole to dipole form. As a result, the
conversion contribution becomes essentially suppressed in comparison to
previous results \cite{nakayam_etapr}.

\section{The reaction $\bf pp \to pp \boldgamma  e^+e^-$}

For the Feynman diagrams in Fig.~\ref{fig0}  generated by the
Lagrangians (\ref{mnn1})-(\ref{em}), the invariant
amplitude $T$ can be cast in a factorized form
\begin{eqnarray}
\label{ampl}
T=T^{(1)}_{NN\to NN\eta'}
\dfrac{i}{q^2-\left(m_{\eta'} -\dfrac{i}{2} \Gamma_{\eta'}\right)^2}
\,T^{(2)}_{\eta'\to\gamma e^+e^-}\, .
\end{eqnarray}
The amplitude $T^{(1)}_{NN\to NN\eta'}$
describes the production process of a pseudoscalar meson (an off-mass shell
$\eta'$ meson in the $NN$ collision), while the  second amplitude
$T^{(2)}_{\eta'\to\gamma e^+e^-}$ describes the Dalitz decay of the produced
meson into a real photon and a di-electron. In the propagator of the $\eta'$ meson
the mass $m_{\eta'}$ is replaced by $m_{\eta'} - i\Gamma_{\eta'}/2$ to take into account
the finite life time of the $\eta'$ meson.

For such factorized Feynman diagrams  one can
separate, in the cross section, the dependence  on the variables connected with
the Dalitz decay vertex and
perform the integration  analytically \cite{ourOmegaFF,ourEta}. Then,
formally, Eq.~(\ref{ampl}) allows one to write the differential cross section (\ref{crossnn})
in a factorized form as well,
\begin{eqnarray}
\frac{d\sigma}{ds_{\eta'} ds_{\gamma^*}} =
\dfrac{d\Gamma_{{\eta'}\to \gamma e^+e^-}}{ds_{\gamma^*}}\,
\frac{1}{4\pi\sqrt{s_{\eta'}}}
\frac{1}{\left(\sqrt{s_{\eta'}}-m_{\eta'}\right)^2 +\frac14 \Gamma_{\eta'}^2}
d^5\sigma^{tot}_{NN\to NN{\eta'}},
\label{two}
\end{eqnarray}
where the integrals over the final di-electron and
photon variables have been carried out analytically  (see  for details
\cite{ourOmegaFF}).
The decay rate $ {d\Gamma}/{d s_{\gamma^*}}$  for the  ${\eta'}$ meson
is defined as
\begin{eqnarray}
\dfrac{d\Gamma_{{\eta'}\to \gamma e^+e^-}}{ds_{\gamma^*}} =
\dfrac{2\alpha_{em}}{3\pi s_{\gamma^*}}
\left( 1-\frac{m_{\eta'}^2}{s_{\gamma^*}}\right)^3
\Gamma_{{\eta'}\to  \gamma \gamma }  \left | F_{ {\eta'}  \gamma \gamma^*\,}(s_{\gamma^*})\right |^2.
\label{dgamma}
\end{eqnarray}
The electromagnetic fine-structure constant is denoted as
$\alpha_{em}$, and $F_{ {\eta'}  \gamma \gamma^*\,}(s_{\gamma^*}) $
is the transition $\eta'\to  \gamma \gamma^*$ form factor.
The  cross section for the production of a pseudoscalar meson with
${\eta'}$ quantum numbers but with  ${s_{\eta'}}\neq m_{\eta'}^2$ is
\begin{eqnarray}
\label{nneta}&&
d^5\sigma^{tot}_{NN\to NN{\eta'}}=\frac{1}{2(2\pi)^5\sqrt{\lambda(s,m^2,m^2)}} \nonumber\\&&
\times \frac14
\sum_{spins}| T_{NN\to NN{\eta'}}^{(1)}|^2 ds_{12}dR_2(N_1N_2\to s_{\eta'} s_{12}) dR_2 (s_{12}\to N_1'N_2'),
\end{eqnarray}
where $s_{12}$ is the invariant mass of the two nucleons
in the final state, and the two-particle invariant phase-space volume $R_2$ reads
\begin{equation}
R_2(ab\to cd)=\frac{\sqrt{\lambda(s_{ab},m_c^2,m_d^2)}}{8s_{ab}} d\Omega^*_c.
\label{ph}
\end{equation}

It can be seen from Eq.~(\ref{two}) that all the peculiarities of the
cross section for the full reaction $NN\to NN \gamma e^+e^-$
are  basically determined by the sub-reaction $NN\to NN{\eta'}$
with creation of a virtual ${\eta'}$ meson. Hence, before
analyzing the full reaction, we shall proceed with a detailed study  of
the sub-reaction  $NN\to NN{\eta'}$, i.e. the
production of an on-mass-shell ${\eta'}$ meson.

\subsection{$\boldeta'$ production in
$\bf NN\to NN{\boldeta'}$ reactions: initial and final state interaction}

It ought to be mentioned that generally the Feynman diagrams deal with
asymptotically free particles, i.e. account
for reactions with non-interacting particles in  initial and final
states. However, in the real process (\ref{dalitzpp}) the two nucleons
can interact in the initial state (ISI) before the ${\eta'}$ creation, and in the final
state (FSI) as well, provoking distortions of the initial and final $NN$ waves.

The initial state distortion due to the $NN$ interaction  before the ${\eta'}$ creation
is to be evaluated at relatively high energies, i.e., larger than the
threshold of the ${\eta'}$ meson production, $T_{kin}\sim 2.41 \, GeV$.
Therefore, one can expect that the variation with the kinetic
energy of ISI effects is small.
As shown in Ref.~\cite{isi}, the effect of ISI
can be factorized in the total cross section and henceforth it plays a role of a
reduction factor in each partial wave in the cross section. This reduction factor
depends  on the inelasticity and  phase shifts of the
partial waves  at the considered energies.
At the threshold,  the number of initial partial wave is strongly limited
by the partial waves of the final states and, in principle,
one can restrict oneself to  $^3P_0$ and  $^1P_1$ waves.
Experimentally  it is found \cite{said} that at kinetic energies of the order of few $GeV$
the phase shifts $^3P_0$ and  $^1P_1$ are indeed almost energy independent and the
reduction factor for each partial wave can be taken constant.
In our calculations  we adopt, for the reduction factor $\zeta$, the expression
from Ref.~\cite{isi}, which
leads to $\zeta=0.277$ for the $^1P_1$ wave and
$\zeta=0.243$ for the $^3P_0$  waves (cf.\ Ref.~\cite{nakayama_eta}).

Final state  interaction  effects in the $NN$ system have been calculated within
the Jost function formalism \cite{gillespe}
which reproduces the singlet and triplet phase shifts at low energies. Details of
calculations of FSI with the Jost function can be found in Ref.~\cite{fsiReznik}.

\subsection{Results for ${\boldeta'}$ meson production in $\bf NN$ collisions}

The amplitude $T_{NN\to NN{\eta'}}^{(1)}$, besides the above listed
meson conversion contribution,
includes the nucleonic and resonance currents, each of them being a 
coherent sum over all the considered exchanged mesons.
The corresponding parameters are described above
in Section \ref{subsecII}.
It is worth mentioning that, in spite of the
large number of  considered diagrams and the large number of effective
parameters, there is not too much freedom in fitting the cross section as long as the known
resonances are considered. However, if one fits data with additional inclusion
of missing resonances, a bulk of parameters
remain practically free and can be fine-tuned to optimize the description of data.
We performed some investigation of such a model with unknown resonances
by considering few missing resonances with masses about $ 2\, GeV/c^2$
and found that, since the energy dependence of the cross section
is rather smooth, various sets of parameters
fit equally well the data near the threshold. 
However, for each of the considered set it turns out
that the cross section rises too rapidly at large values of the excess energy.
Therefore, to reconcile this  dependence with a smoother behavior
at large excess energies one can, in principle, add more
and more resonances and adjust appropriately the relative (unknown)
signs of the couplings
to compensate contributions from different diagrams
(the interested reader can find a detailed
analysis of such a situation in Ref.~\cite{nakayam_consist}).

\begin{figure}[h]  
\includegraphics[width=1.0\textwidth]{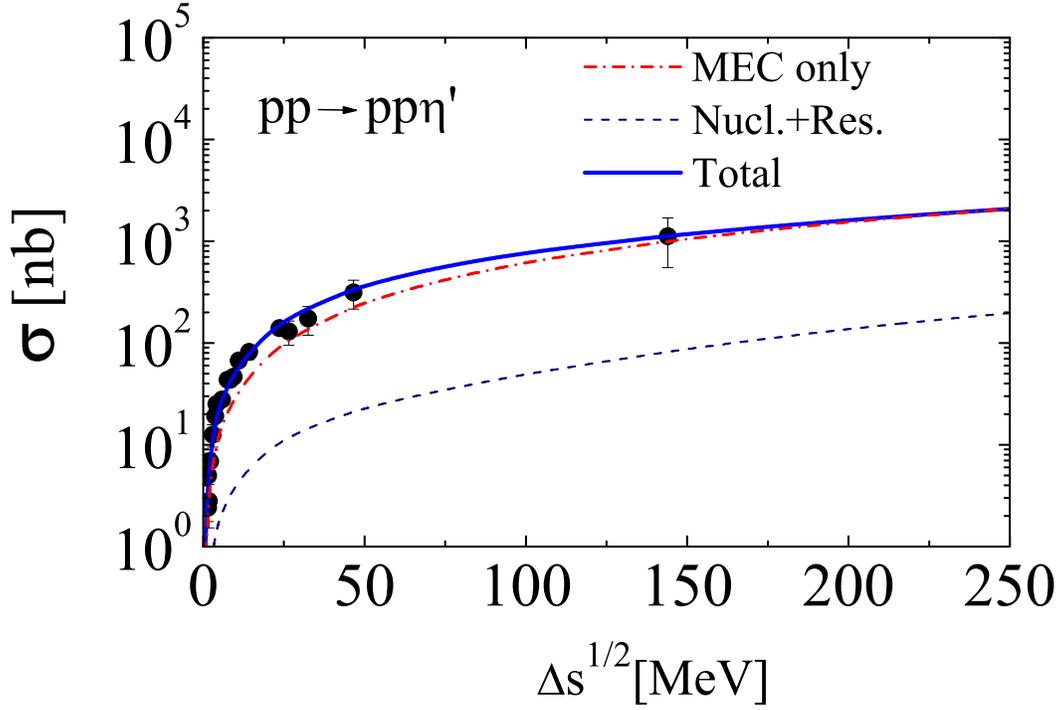}
\caption{Total  cross section for the $pp\to pp{\eta'}$ reaction
as a function of the excess energy $\Delta s^{ 1/2}=(\sqrt{s}-2m-m_{\eta'})$.
The dot-dashed line depicts the mesonic conversion
contribution (MEC only), while the dashed curve represents the total contribution
of nucleonic and resonances currents (Nucl.+Res.). 
The solid curve is the total contribution.
Data are from Ref.~\cite{experEtaprime}.}
\label{fig1}
\end{figure}

\begin{figure}[h]  
\includegraphics[width=1.0\textwidth]{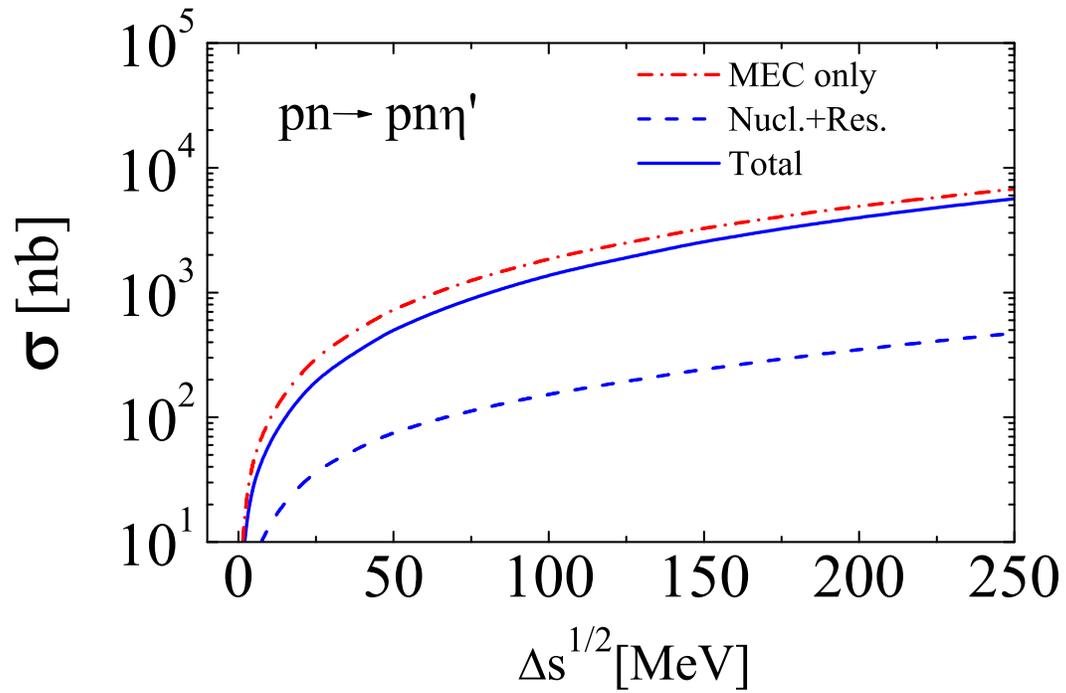}
\caption{The same as in Fig.~\ref{fig2} but for the $pn\to pn{\eta'}$ reaction.}
\label{fig2}
\end{figure}

In our present calculations we included
only the known  four-stars resonances $S_{11}(1650)$, $P_{11}(1710)$ and  $P_{13}(1720)$.
Results of numerical  calculations for the total cross section
$\sigma^{NN\to NN{\eta'}}$  are presented
in Figs.~\ref{fig1} and~\ref{fig2},
for proton-proton and proton-neutron reactions, respectively.
Experimental data are taken from \cite{experEtaprime}.
The total contribution of   nucleonic and resonance currents
(dot-dashed lines) is found to
be much smaller than the mesonic diagrams alone  providing a constructive  interference
in $pp$ and a destructive one in $pn$ reactions. As already discussed, the small contribution
of the nucleonic
and resonance currents is due to the smallness of the  coupling constants $g_{\eta'NN}$ and
$g_{\eta'N^*N}$. The magnitude of other parameters entering in our calculations
are dictated by the known experimental data.
A reasonable agreement with data is achieved almost by the contribution of the
meson conversion currents. This does not contradict to the results reported in
Ref.~\cite{nakayam_etapr}, however, if one adopts a dipole form factor \cite{nakayam_consist}
for the meson conversion vertex, then additional resonances with freely adjustable parameters
have to be accommodated in  the model.

\begin{figure}[h]  
\includegraphics[width=0.9\textwidth, height=0.9\textheight]{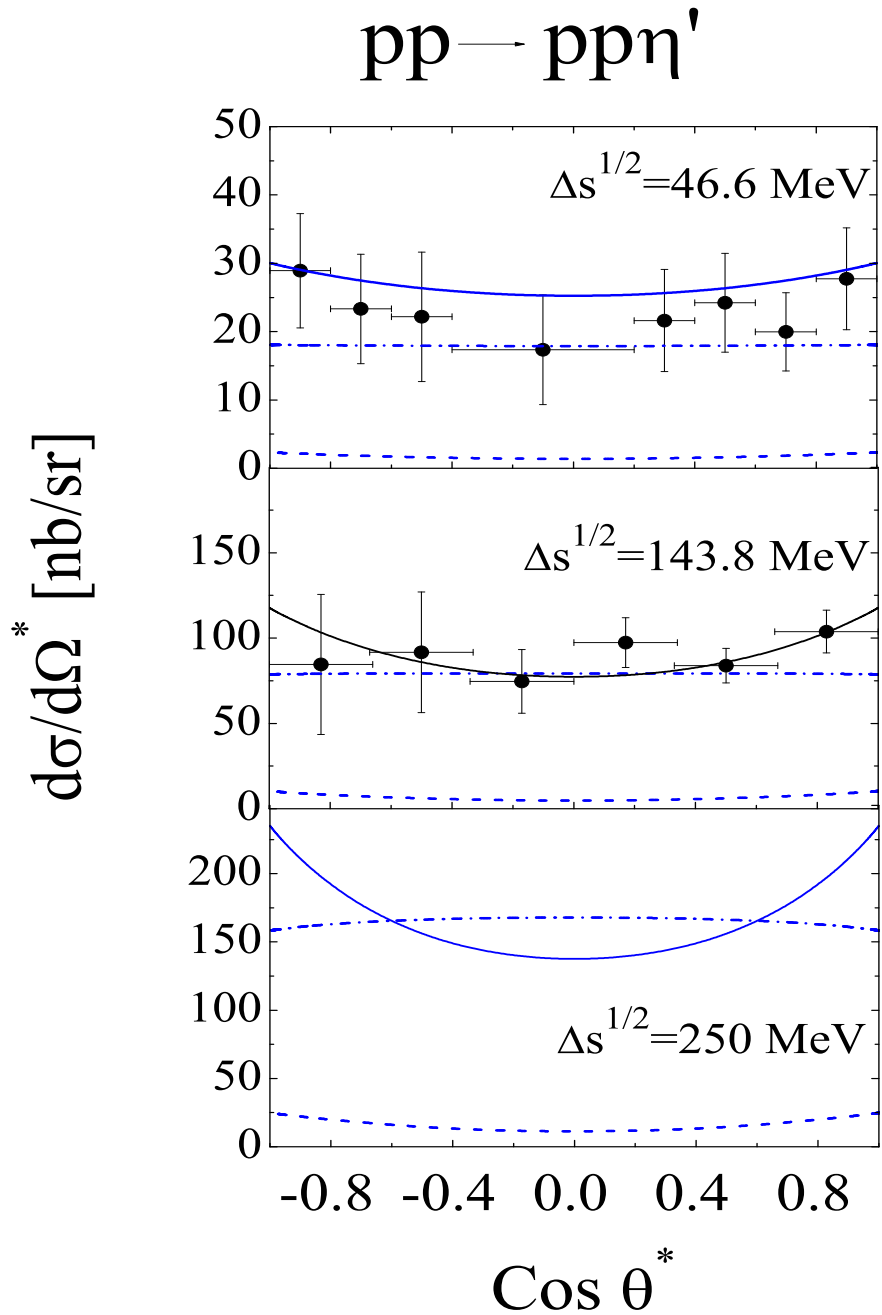} %
\caption{Angular distribution of ${\eta'}$ mesons in the center-of-mass system
in $pp\to pp\eta'$ reactions at three different excess energies. 
Line codes as in Fig.~\ref{fig1}.
Data are from Ref.~\cite{experEtaprime}.}
\label{fig3}
\end{figure}

\begin{figure}[h]  
\includegraphics[width=1.0\textwidth]{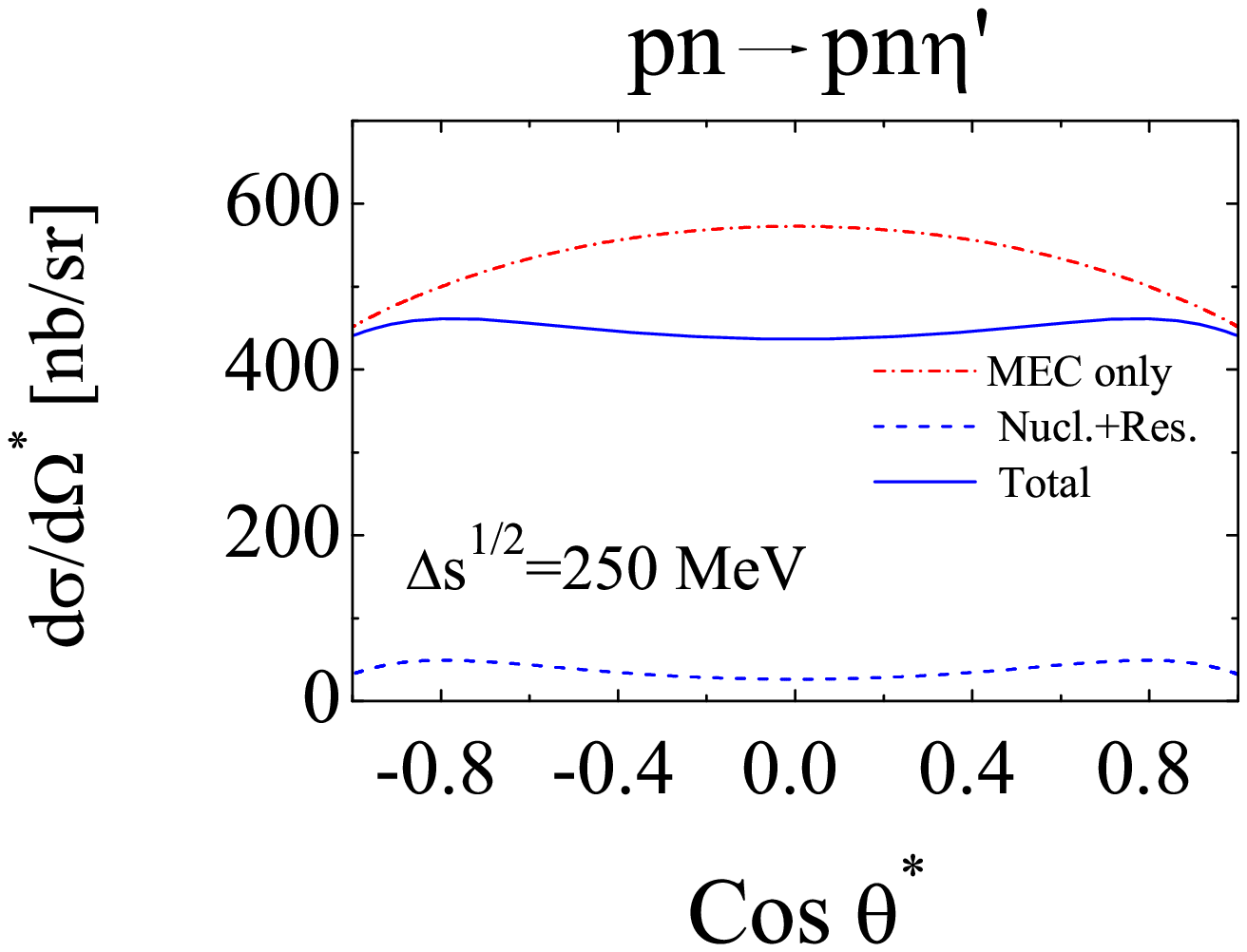} %
\caption{Angular distribution of ${\eta'}$ meson in the
center-of-mass system
in $pn\to pn\eta'$ reactions.  
Line codes as in Fig.~\ref{fig1} }
\label{fig4}
\end{figure}

To further investigate the role of
different diagrams, one needs to consider quantities being more sensible
to the production mechanism. For instance, a detailed investigation of the angular distribution
(see Figs.~\ref{fig3} and \ref{fig4}) shows that the mesonic current provides a
form of the differential cross section which evolves  from
a flat to a convex curve with increasing excess energy.
Contrarily, the nucleonic  and resonance currents give a
concave shape of the corresponding contributions. The interference effects (which
are maximum
in forward-backward directions) lead to a slightly convex form of the resulting curves.
Experimental data are still too scarce to determine more precisely the relative
contributions of different diagrams. Note again that, taking into account
more resonances, the description of the experimental data
on angular distribution at larger excess energies cannot be
improved \cite{nakayam_consist}. New data will essentially enlighten the problem.

Note also that, even achieving a good fit of the cross section in proton-proton
reactions, it is not \textit{a priory}
clear whether the obtained set of parameters equally well describes also the proton-neutron
reactions. The isospin dependence of the amplitude is determined by a subtle
interplay of different diagrams with different exchange mesons (scalar, vector,
iso-scalar, iso-vector etc.). Once the parameters for $pp$ reactions
are fixed, the $pn$ amplitude  follows directly from this  set of parameters without any
further readjustment. Of course,
the ISI and FSI factors are different for $pp$ and
$pn$ systems, but they are fixed by independent information.
Our results displayed in Fig.~\ref{fig2}-\ref{fig4}  may
serve as predictions for the $pn$ channel.

\begin{figure}[h]  
\includegraphics[width=1.0\textwidth]{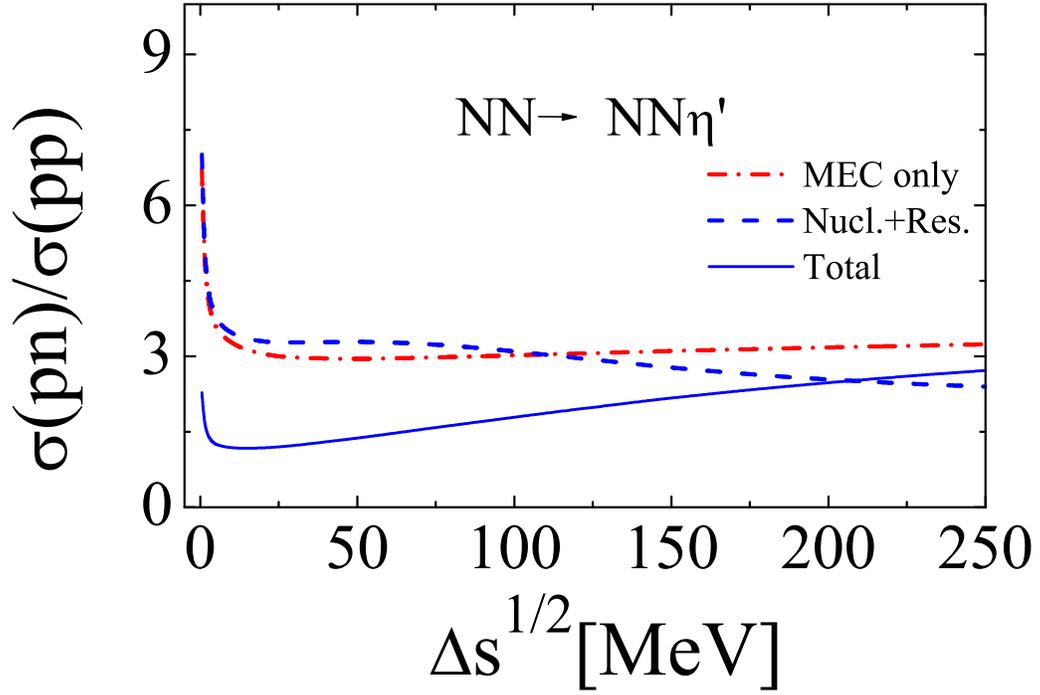} %
\caption{ Ratio of the cross sections $pn\to pn\eta'$
to $pp\to pp\eta'$ as a function of the excess energy. 
Line codes as in Fig.~\ref{fig1}.}
\label{fig5}
\end{figure}

Figure \ref{fig5} exhibits the isospin dependence of the
total cross section. Similar to $\omega$, $\phi$ and $\eta$ production
\cite{ourOmegaFF,ourOmega}
the ratio of the $pn$ channel to the $pp$ channel depends
on the excess energy. This means that a simple (i.e.\ constant)
isospin factor (see discussions in Ref.~\cite{hanhart}) cannot relate these channels.
In addition, our calculations point also to the possibility of rather different 
angular distributions
at the same excess energy, as evidenced in Figs.~\ref{fig3} and \ref{fig4}. 
The angular distribution
in the $pn$ channel is fairly flat. The rise of the isospin dependence at low
excess energies is entirely determined by the FSI effects which are rather different 
in the $pp$ and
$pn$ channels. At larger excess energy the FSI effects vanish and the isospin  ratio
is determined entirely by the corresponding Feynman diagrams.   
Figure~\ref{fig5} exposes the importance
of the $pn$ channel for heavy-ion collisions. This seems to hold for the production of light
vector mesons and virtual ($e^+e^-$) bremsstrahlung too \cite{ourNuclPhys}.

\section{Daltz decay}

\subsection{Preliminaries}

Consider first the general process of Dalitz decay of an on-mass shell pseudo-scalar 
meson $P$ into a real and a virtual photon (di-electron). Within
the present approach this reaction can be considered as a
two-stage process, when the meson decays firstly into two photons, and
secondly, one of the virtual photon decays into a di-electron pair.
The decay width of production of two real photons
is calculated from Eq.~(\ref{em}) as
\begin{eqnarray}
\Gamma_{P\to  \gamma \gamma}=
\frac{s_{P}^{3/2}}{64\pi}
f_{ P\,\gamma \gamma }^2(0)
\label{g0}
\end{eqnarray}
and serves for a determination of the coupling constant $f_{ P\,\gamma \gamma }$
from experimental data.
The square of the $\gamma \gamma$ invariant mass is denoted by $s_P$. Since in what follows
we are interested in production and decay of ${\eta'}$ mesons, being generally
off-mass shell,  we use the notation $s_{\eta'}$  instead of $s_P$, bearing in
mind that $s_{\eta'}\neq m_{\eta'}^2$.
Contrarily to the vector meson case \cite{ourOmegaFF},
instead of the factor $1/3$ (due to averaging over three projections of the
spin of the vector particle) in Eq.~(\ref{g0}) a factor of $1/2$ appears due to
two photons in the final state.
Equation~(\ref{g0}) yields
$|f_{{\eta'}  \gamma \gamma\,}|\simeq 0.0314\ \ GeV^{-1}$
for the known  width $\Gamma_{\eta'\to\gamma\gamma}=( 4.30\pm 0.15)\  keV$ \cite{pdg}.
In our  calculations   the sign of the coupling constant
has been taken positively.

The transition form factor
$F_{ {\eta'}  \gamma \gamma^*\,}(s_{\gamma^*}) $
requires separate consideration. As stressed above,
the functional dependence
of form factors upon the momentum transfer is a source of information on
general characteristics of hadrons, such as  charge and magnetic
distributions, size etc. Also, form factors are known as important quantities characterizing
bound states within non-perturbative QCD.
The main theoretical tools in studying  exclusive processes
within the non-perturbative QCD  are approaches based on light cone sum rules
and factorization theorem (see, e.g., \cite{rad1,rad,novikov,isgur,isgur1}
and references therein  quoted). At high $Q^2$, i.e., at high virtualities of quarks
the hard gluon exchange is predicted to be dominant. However, since the virtuality
of a quark is determined by $x_i p$ ($x_i$ is fraction part of the total momentum $p$
carried by quark) the QCD asymptotic regime for low values of $x_i$
can be postponed until extremely large
values of the momentum transfer and non-perturbative (soft) contributions
play an important role in exclusive processes. In this connection an
investigation of form factors in large intervals of momentum transfer,
including the time-like region, serves as an
important tool to provide additional information about the
QCD regimes and the  interplay between soft and hard
contributions.  Besides, there is a number of more phenomenological models, e.g., based on
the dispersion relation technique \cite{anisovich_IE,omegaFormfNormaliz}, or  on
use of vector meson dominance 
models (for details see Refs.~\cite{faessler,thomas,rmp,meissner}), or on
effective SU(3) chiral Lagrangians with inclusion of the $U(1)$ 
non-Abelian anomaly \cite{meissner,kaiser,nonAbelian}.
It is also known  that the asymptotic  behavior of the transition form factors can be
determined from the corresponding axial anomaly \cite{adler}.
Traditionally,  the electromagnetic form factors are studied by electron scattering
from stable particles which provides information  in the space-like region of momenta
where, as well known, the experimental data can be peerless parameterized by dipole formulae.
This in turn means that in the unphysical region, i.e., for kinematics
outside the  reach of experiments with on-mass shell particles,
the analytically continued  form factors display a pole structure.
Intensively studied form factors
are the ones of the light pseudo-scalar mesons, in particular the pion.
Heavier pseudo-scalar meson form factors have received less attention since their experimental
determination is more difficult. However, there are already experimental
data on form factors of the heavier pseudoscalar mesons, such as
$\eta$ and $\eta'$, most of them  in
the space-like region \cite{CLEO_FF,transFF_exper}.
With the spectrometer HADES \cite{HADES}  which can investigate
with a high efficiency di-electrons production in $pp$ collisions
in a wide kinematical range of invariant masses, an additional
study of the transition form factor for $\eta' \to \gamma e^+e^-$ process becomes
feasible.

The simplest and quite successful
theoretical description of form factors can be performed \cite{faessler,rmp,meissner}
with the VMD conjecture,
within which a reasonably good description of elastic form factors in the time-like region
has been accomplished.
By using the current-field  identity~\cite{meissner,faessler}
\begin{equation}
J^\mu = -e\frac{m_\rho^2}{f_{\gamma\rho}}\Phi_{\rho^0}^\mu
-e\frac{m_\omega^2}{f_{\gamma\omega}}\Phi_{\omega}^\mu
-e\frac{m_\phi^2}{f_{\gamma\phi}}\Phi_{\phi}^\mu +\cdots
\label{currid}
\end{equation}
with the coupling constants $f_{\gamma\rho}$,  $ f_{\gamma\omega}$ and $f_{\gamma\phi}$
known \cite{feldmann_PSmesons,faessler,friman1,friman2}
from experimentally measured electromagnetic decay widths,
one can  also  compute the transition form
factor  $F^{VMD}_{ {\eta'} \gamma \gamma^*\,}(s_{\gamma^*})$ by
evaluating the corresponding Feynman diagrams.  The result is
\begin{eqnarray}&&
F^{VMD}_{ {\eta'} \gamma \gamma^*\,}(s_{\gamma^*})=
\sqrt{4\pi\alpha_{em}}
\sum\limits_{V=\rho,\omega,\phi...}
\frac{f_{\eta' V \gamma}}{f_{\eta'\gamma\gamma} f_{\gamma V}}
\frac{m_V^2}{m_V^2-s_{\gamma^*}}\equiv
\sum_{V=\rho,\omega,\phi...}C_V\frac{m_V^2}{m_V^2-s_{\gamma^*}},
\label{fvmd}
\end{eqnarray}
where the summation over vector mesons $V$, besides $\rho,\omega$ and $\phi$ mesons, 
contains also  heavier
vector particles~\cite{faessler}. The normalization of the form factor
at the origin, $F_{\eta' \gamma \gamma^* }(s_{\gamma^*}=0)=1$ implies that
$\sum\limits_V C_V = 1$. Actually, in real calculations one takes into account
only a restricted number of vector mesons, for which  $\sum\limits_V C_V \neq 1$,
i.e. in to obtain correct normalization, the coefficients $C_V$ 
must be rescaled.
Here,we take into account $\rho, \,\omega$ and $\phi$ mesons.
The absolute values  of  $C_V$ in Eq.~(\ref{fvmd})
have been calculated from the rates Eq.~(\ref{g0}) as
\begin{equation}
\label{rates}
\Gamma(V\to e^+e^-)=\frac{4\pi\alpha_{em}^2}{3}\frac{m_V}{f_{\gamma V}^2}.
\end{equation}
The relative signs
have to be  fixed from some more general considerations. 
For instance, SU(3) symmetry
implies an opposite sign for $f_{\gamma\phi}$ relative to the
corresponding couplings for $\rho$ and $\omega$, viz.
$f_{\gamma\rho}:f_{\gamma\omega}:f_{\gamma\phi}=1:3:(-3/\sqrt{2})$.
We suppose that the sign  of the coupling constant  $f_{\eta' V \gamma}$ does
not depend upon the  mass of the vector particle, i.e., the ratio
 $f_{\eta' V \gamma}/f_{\eta' \gamma\gamma} $
in the present paper is taken positive for all mesons.
Numerical calculations of $C_V$ with experimental decay rates \cite{pdg}
result in $C_\rho=0.946$, $C_\omega=0.095$ and $C_\phi=-0.041$, with a rescaling factor
$\sim 0.83$.
The accordingly computed
transition form factors (see also Refs.~\cite{transFF_VMD,feldmann_FF,LandPhysRep})
are in a reasonable agreement with the available experimental
data \cite{transFF_exper}, which usually are parameterized in
a monopole form
\begin{equation}
F_{\eta' \gamma \gamma^* }(Q^2) =
\frac{1}{1-Q^2/\Lambda_{\eta'}^2},
\label{monopole}
\end{equation}
where the effective pole-mass parameter
$\Lambda_{\eta'}\simeq 0.81 GeV$ is a combination of the masses of the
vector mesons. It is worth emphasizing  that the experimental
parameter $\Lambda_{\eta'}$ has been obtained by fitting
data mostly in the space-like region ($Q^2 <0$).
In case of time-like arguments of the form factors,
i.e., when $Q^2\equiv s_{\gamma^*} > 0$ the pole mass parameters $m_V$ in
Eq.~(\ref{fvmd})
receive a small imaginary part, proportional to the total $V$ meson widths,
$m_V\to m_V - i\Gamma_V/2$.
For the monopole parameter $\Lambda_{\eta'}$ in Eq.~(\ref{monopole}),
we adopt the imaginary part as $ \Gamma_\rho/3$.

\begin{figure}[h]  
\includegraphics[width=1.0\textwidth]{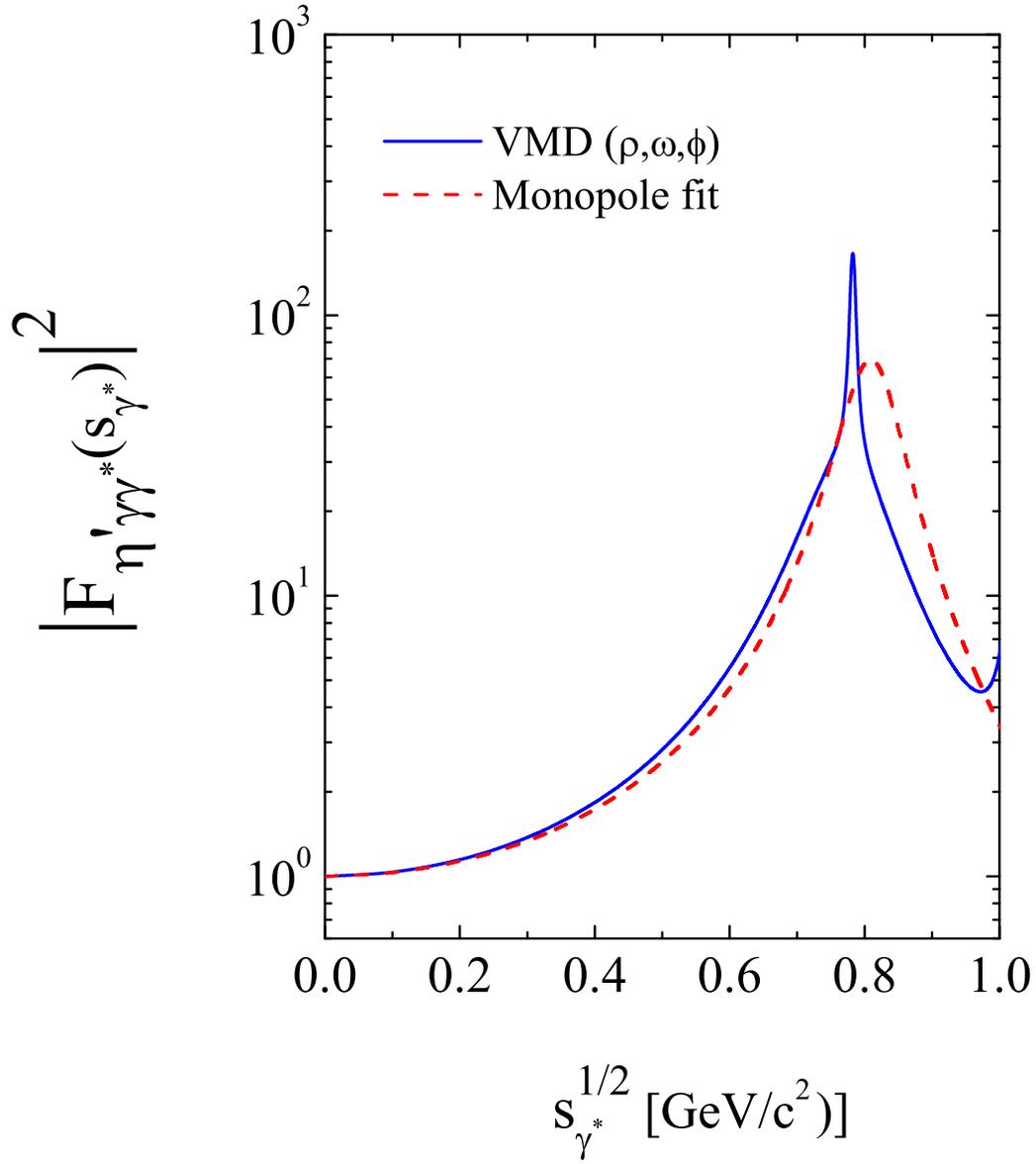} %
\caption{Transition form factor $\eta'\to\gamma\gamma^*$
computed within the VMD model (solid line) in comparison with
the  monopole fit (dashed line)
$F_{\eta'\gamma\gamma^*} =1/(1-{s_\gamma^*}/\Lambda_{\eta'}^2)$
with $\Lambda_{\eta'}=(0.81-i\Gamma_\rho/3$) \cite{CLEO_FF,transFF_exper}.}
\label{fig6}
\end{figure}

In Fig.~\ref{fig6} we compare the transition form factor
computed by the VMD formula, Eq.~(\ref{fvmd}),
with the corresponding monopole experimental fit (\ref{monopole}). It is seen that the
form factors obtained within VMD model and fitted to  data
are in a good agreement in a wide interval of di-electron masses.
Note that in the Dalitz decay $\eta'\to \gamma e^+e^-$ the $\rho$ and $\omega$
poles occur in the physical region allowed for the invariant mass of
the di-electrons. This implies that an experimental investigation of
the $e^+e^-$ mass spectrum in the vicinity of the $\rho$ and $\omega$
poles will provide an excellent test of the validity of the VMD conjecture
at large virtualities in the time-like region.

\section{The complete $\bf NN \to NN \gamma e^+e^-$  reaction}

In order to emphasize the dependence upon the di-electron invariant mass  we
rewrite the cross section (\ref{two}) in the form
\begin{eqnarray}&&
\frac{d\sigma}{ds_{\gamma^*}}=\frac{2\alpha_{em}}{3\pi^2}
\left|F_{\eta' \gamma \gamma^* }(s_{\gamma^*})\right|^2
\frac{Br({\eta'}\to 2\gamma)}{m_{\eta'}^2}{\cal I}(s_{\gamma^*}),\nonumber\\&&
{\cal I}(s_{\gamma^*})\equiv\int\limits_1^{\xi_{max}}
d \xi \frac{(\xi-1)^3}{\xi}
\frac{b/a}{(\xi-a)^2+b^2}\sigma^{tot}_{NN\to NN{\eta'}}(\xi),
\label{xi}
\end{eqnarray}
where we introduced dimensionless variables $a$,$b$ and $\xi$ as follows
\begin{equation}
a\equiv \frac{m_{\eta'}^2}{s_{\gamma^*}};\qquad b\equiv \frac{m_{\eta'}\, \Gamma_{\eta'}}{s_{\gamma^*}};
\qquad \xi\equiv \frac{s_{\eta'}}{s_{\gamma^*}}; \quad \xi_{max}=\frac{(\sqrt{s}-2m)^2}{s_{\gamma^*}}
\label{pae}
\end{equation}
and $Br({\eta'}\to 2\gamma)$ is the branching
ratio of the ${\eta'}$ meson decay into two real photons,
$Br({\eta'}\to 2\gamma)=(2.12\pm0.14)\%$ \cite{pdg}.
Since the total width of the ${\eta'}$  meson
is fairly small, the parameter $b$ in (\ref{xi}) provides a
sharp maximum of the integrand function at $\xi=a$ as far as
the parameter $a>1$ (or, equivalently, the di-electron invariant mass
$s_\gamma^* < m_{\eta'}^2$). This allows to take out from the integral the smooth
function $\sigma^{tot}_{NN\to NN{\eta'}}(\xi=a)$ and calculate it at
$s_{\eta'}=m_{\eta'}^2$, i.e. at the ${\eta'}$ meson pole  mass. The remaining
integral can then
be carried out analytically. However, at $a<1$ when the di-electron mass is larger
than the $\eta'$ pole mass, i.e., the integrand does not exhibit anymore a resonant shape, and
the integral needs to be calculated numerically.

Equation~(\ref{xi}) shows that the di-electron invariant mass distribution
$ {d\sigma}/{ds_{\gamma^*}}$ is  proportional to the transition form factor
$F_{{\eta'}\gamma\gamma^*}(s_{\gamma^*})$ so that measurements of this distribution
provide a direct experimental information about this quantity.
At values $s_{\gamma^*} < m_{\eta'}^2$ the smooth part of the
integrand, $\dfrac{b(\xi-1)^3}{a\xi}\,\sigma^{NN\to NN{\eta'}}(\xi)$, can
be pulled out  the integral at $\xi=a$ obtaining
\begin{eqnarray}
\label{approxint}
{\cal I}(s_{\gamma^*})=
\dfrac{\pi(a-1)^3}{a^2}\,\sigma^{NN\to NN{\eta'}} (s_{\eta'}=m_{\eta'}^2).
\end{eqnarray}
Equations (\ref{xi}) and (\ref{approxint}) allow to define an experimentally
measurable ratio which is directly proportional to the form factor
\begin{eqnarray}
\label{epsFF}
R(s_{\gamma^*})=
\frac{d\sigma/ds_{\gamma^*}}{\left.
\left (d\sigma/ds_{\gamma^*}\right)\right|_{s_{\gamma^*}=s_{\gamma^* \,min}}}
\frac{s_{\gamma^*} }{s_{\gamma^*\, min}}
\left( \frac{1-s_{\gamma^*\, min}/m_{\eta'}^2}{1-s_{\gamma^*}/m_{\eta'}^2}\right)^3=
\frac{|F_{{\eta'}\gamma\gamma^*}( s_{\gamma^*})|^2}{|F_{{\eta'}\gamma\gamma^*}( s_{\gamma^* \, min})|^2},
\end{eqnarray}
where $s_{\gamma^* \,min} $ is the minimum  value accessible experimentally
(in the ideal case  this is the kinematical limit  $s_{\gamma^* \,min}=4\mu_e^2 $).  At low
enough values of $s_{\gamma^* \,min} $, the transition form factor  is close to its normalization
point $F_{{\eta'}\gamma\gamma^*}(0)=1$ and the ratio (\ref{epsFF}) is just the
transition form factor as a function of $s_{\gamma^*}$.

As $a$ approaches unity keeping the maximum position  still within the integration range, 
one can again take out the integral the smooth function 
$\sigma^{NN\to NN{\eta'}}$ at $s_{\eta'}=m_{\eta'}^2$.
However, now the function $(\xi-1)^3/\xi$ cannot be considered smooth enough
and must be kept under the integration. Nevertheless, even
in this case the remaining integral can be computed analytically \cite{ourEta}
and one can still define a ratio analogously to Eq.~(\ref{epsFF})
which allows for an experimental
investigation of the form factor  near the free ${\eta'}$ threshold 
($s_{\gamma^*}\to m_{\eta'}^2$).
At $s_{\gamma^*}> m_{\eta'}^2$ the integral ${\cal I}(s_{\gamma^*})$ 
does not have anymore a sharp
maximum and it ought to be calculated numerically.

\begin{figure}[h]  
\includegraphics[width=1.0\textwidth]{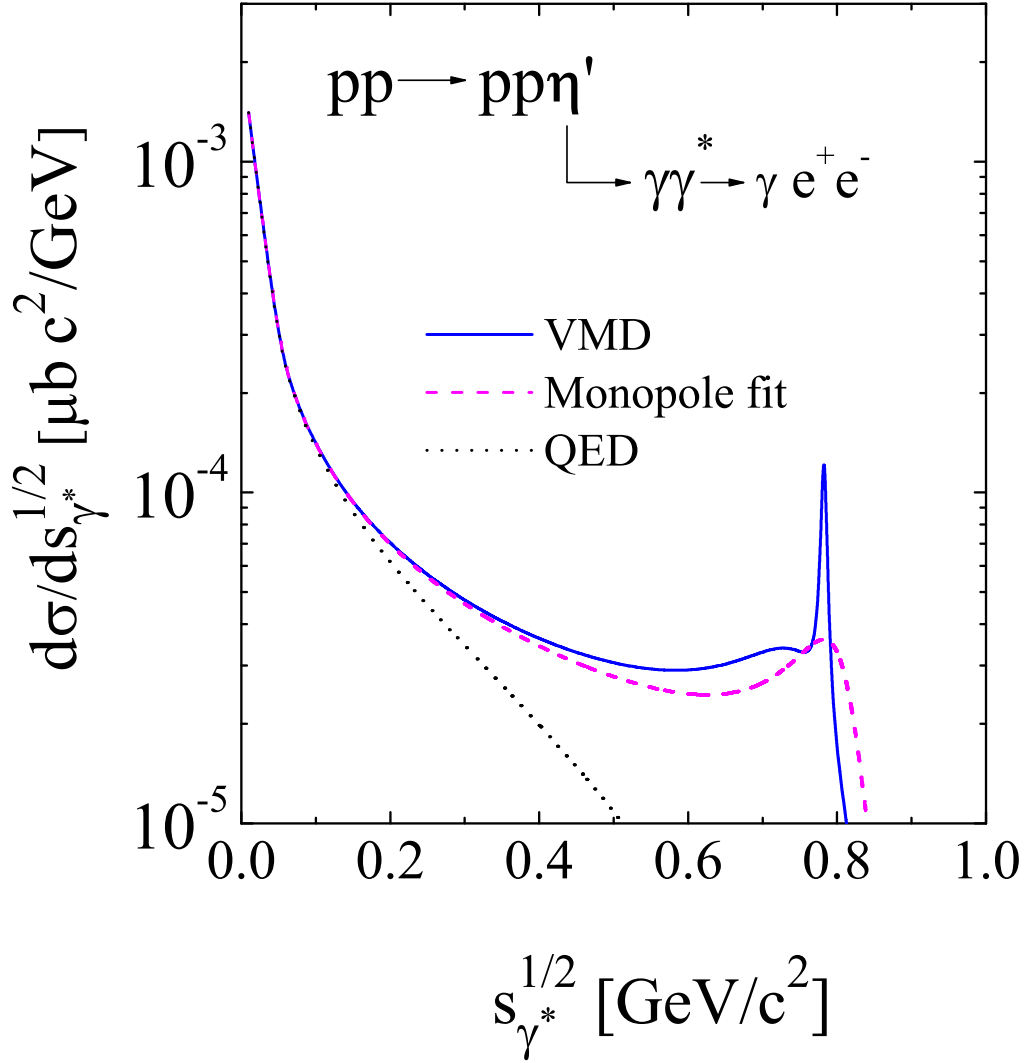} %
\caption{Invariant mass distribution of di-electrons
from   $\eta'$ Dalitz decay in the reaction
$pp= pp \eta'\to pp \gamma e^+e^-$
at proton kinetic energy $T_p=2.5\,  GeV$ (corresponding
to the excess energy $\Delta s^{1/2}\simeq 31.5\,  MeV$).
Solid line - calculations with the transition form factor
within the vector meson dominance model,
dashed line - transition form factor fit from Fig.~\ref{fig6}, 
dotted line -  pure QED form factor,
meaning $F_{\eta'\gamma\gamma^*}=1$. }
\label{fig7}
\end{figure}

\begin{figure}[h]  
\includegraphics[width=1.0\textwidth]{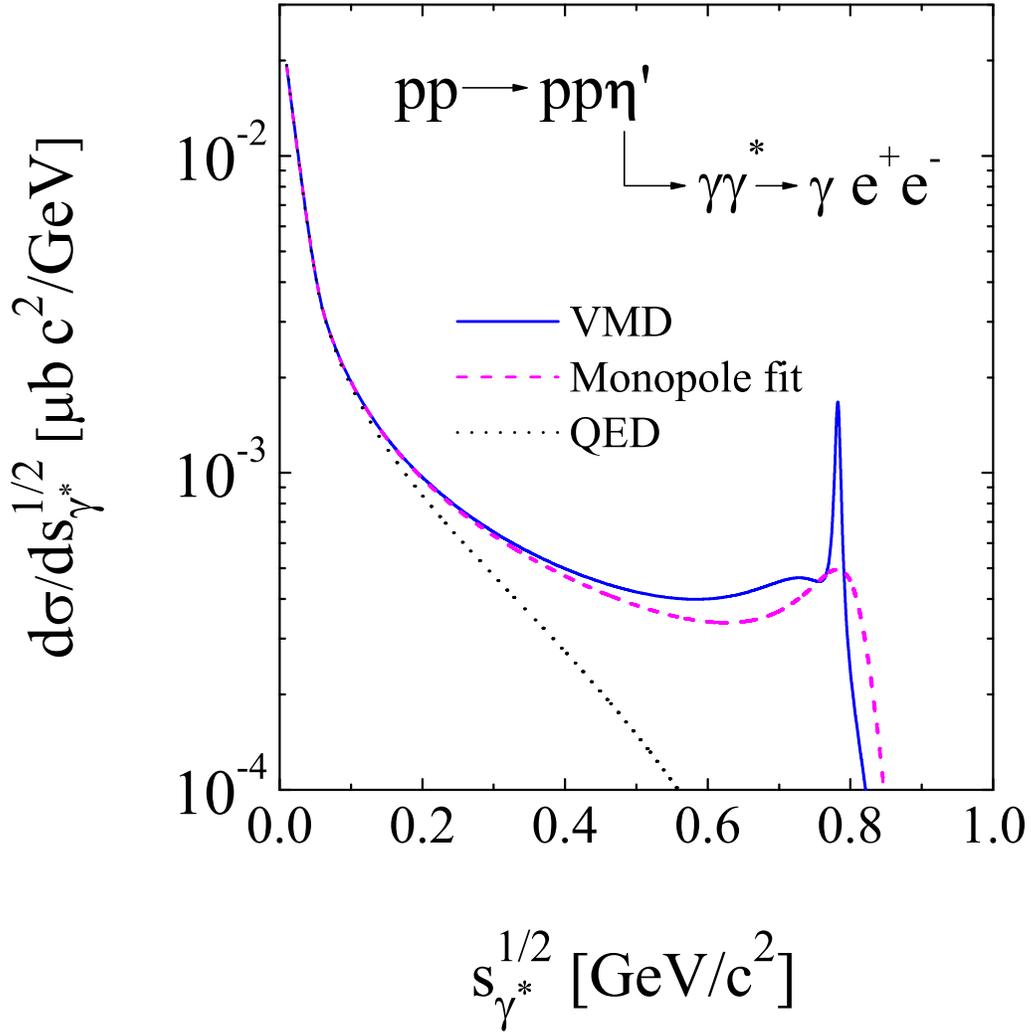} %
\caption{Same as in Fig. \ref{fig7}, but at $T_{kin}=3.5\,  GeV$
corresponding to   $\Delta s^{1/2}\simeq 342\ MeV$.}
\label{fig8}
\end{figure}

In Figs.~\ref{fig7} and~\ref{fig8}  the di-electron invariant mass distribution
${d\sigma}/{d {s^{1/2}_{\gamma^*}}}$
is exhibited as a function of the invariant mass $\sqrt{s_{\gamma^*}}$ calculated by
(i) Eq.~(\ref{xi}) with the transition form factor
$F_{{\eta'}\gamma\gamma^*}(s_{\gamma^*})$ from
the VMD model (solid line), 
(ii) with the monopole experimental fit
(\ref{monopole}) (dashed line), and 
(iii) without accounting for the
form factor, i.e.\ with $F_{{\eta'}\gamma\gamma^*}(s_{\gamma^*})=1$, known also as
QED calculations (dotted line).
It is clearly seen that without form factors the calculated cross section
rapidly vanishes as $\sqrt{s_{\gamma^*}}\to m_{\eta'}$,
while the inclusion of the form factors results in a resonant shape of the distribution.

\begin{figure}[h]  
\includegraphics[width=1.0\textwidth]{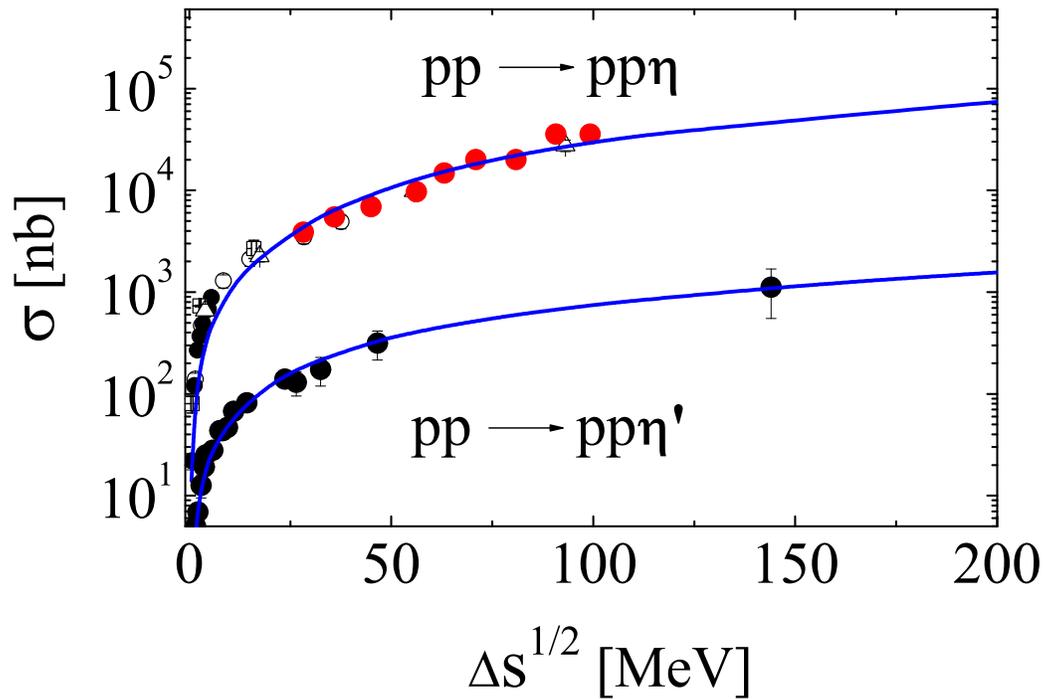}
\caption{Comparison of total cross sections for ${\eta}$ and ${\eta'}$
production in $pp$ collisions as a function of the excess energy.
Details of calculations and references for data of the $\eta$ meson
cross section can be found in Ref.~\cite{ourEta}.}
\label{fig1a}
\end{figure}

It should be noted that the considered final state
(two nucleons, one real photon and one di-electron) can be produced by
various competing channels too. For instance, prominent channels are
Dalitz decays of pions and $\eta$ mesons as well. However, the contribution of
different pseudo-scalar mesons to the di-electron mass
distribution is well separated  kinematically. Thus, at low values of $s_{\gamma^*}$
the pion contribution is predominant \cite{barz}. At intermediate
$s_{\gamma^*}$, close to the $\eta$ meson pole mass, the invariant
contribution of $\eta$ Dalitz decay is much larger than the corresponding
contribution from $\eta'$ mesons. This can be understood
from an inspection of Eq.~(\ref{xi}) which is valid also for
$\eta$  meson  production. The main difference in mass distributions
occurs from different values of the total cross section $\sigma^{tot}_{NN\to NNP}$
for $\eta$ and  $\eta'$ production. For an illustration,
in Fig.~\ref{fig1a} we present a comparison of
the total $\eta$ and $\eta'$ cross sections as a function
of the excess energy $\Delta s^{1/2}$.
It is seen that the shape of the energy dependence is basically the same, however, the
$\eta$ production cross section, at equal excess energy,
is larger by more than one order of magnitude.
Moreover, at equal initial kinetic beam energy the $\eta$ curve is essentially
shifted towards higher excess energies. For instance, at $T=2.5 \, GeV$ the
excess energy for $\eta'$ is $\Delta s^{1/2} \sim 32 \, MeV$ while for $\eta$ meson
$\Delta s^{1/2}\sim 442\, MeV$. In this case,
the $\eta$ cross section becomes larger than the $\eta'$
cross section by more than  three  orders of magnitude.
The remaining terms in Eq.~(\ref{xi}) are of the same order for both
$\eta$ and $\eta'$ mesons.
Hence, at $s_{\gamma^*}$  close to the $\eta$ pole mass the invariant mass distribution,
for the specified final state with one real photon and one di-electron, is entirely
governed by the $\eta$ Dalitz decay.  However, at larger di-electron masses,
where the (virtual) $\eta$ meson is highly
off-mass shell, this contribution rapidly vanishes \cite{ourEta} and,
consequently, beyond  the $\eta$ pole mass the distribution
is governed by the $\eta'$ Dalitz decay.

\begin{figure}[h]  
\includegraphics[width=1.0\textwidth]{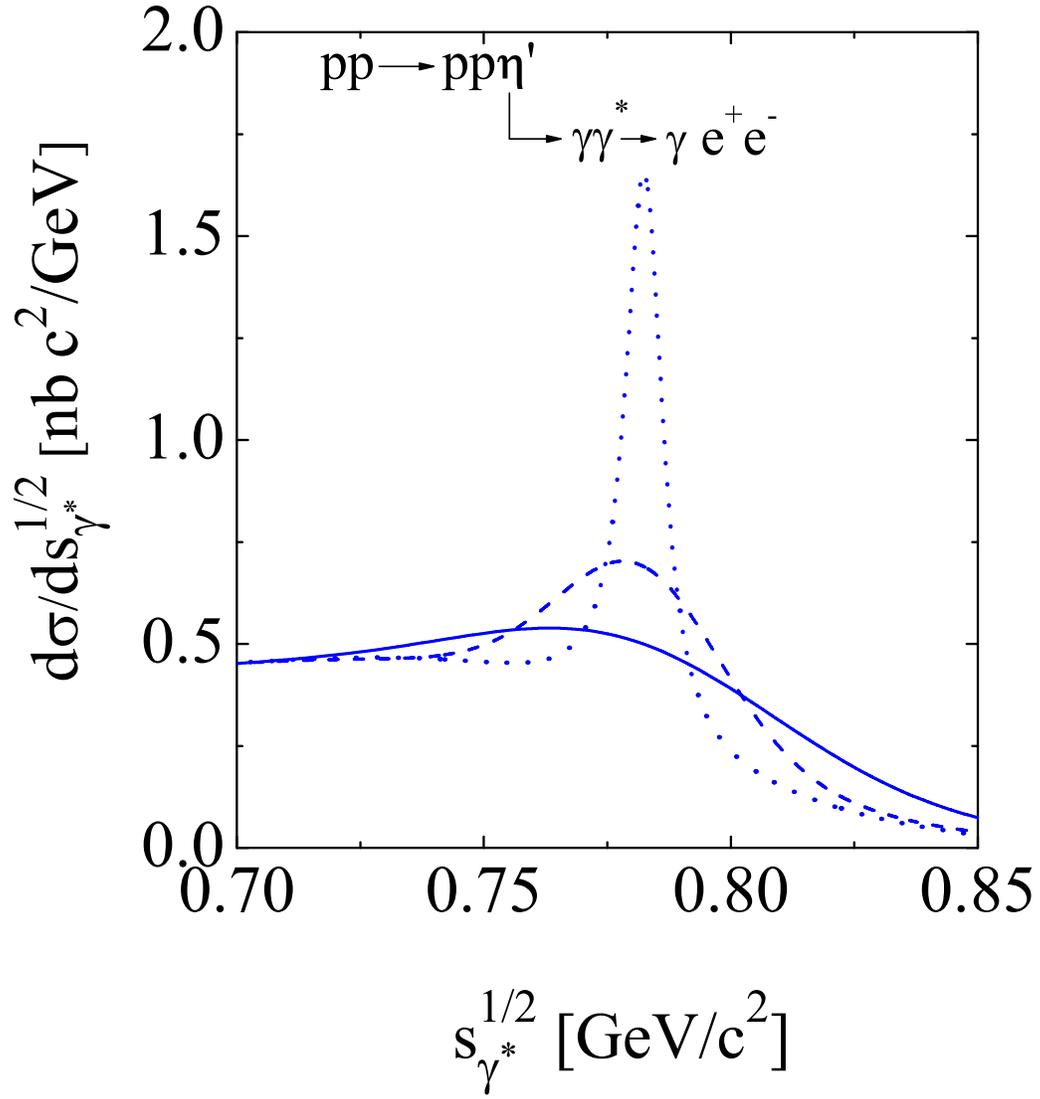}
\caption{ Invariant mass distribution weighted with a Gaussian
corresponding to a finite mass resolution of the detector.
Solid (dashed) line corresponds to the dispersion
$d=30\, MeV$ ($d=15\, MeV$), while dotted line is without smearing.
For $T_{kin} = 3.5$ GeV.}
\label{fig10}
\end{figure}

Nevertheless, some comments about this kinematical  region are in order. Since the
$\eta'$ mass is above the $\rho$ and $\omega$ pole masses, the transition form factor and,
correspondingly the mass distribution, display a sharp peak due to smallness of
the $\omega$ meson width. To get a feeling on the required detector resolution
to identify experimentally such a peak, we simulate it by
\begin{equation}
\label{smear}
\frac{d\sigma}{ds^{1/2}_{\gamma^*}}
={\cal N} \int \limits_{\sqrt{s_{\gamma^*}}-4d}^{\sqrt{s_{\gamma^*}}+4d}
\frac{d\sigma}{d\tilde m } \exp{\left(-\frac{(\tilde m -s^{1/2}_{\gamma^*} )^2}{2d^2}\right)}
d\tilde m                     .
\end{equation}
In Fig.~\ref{fig10}
we present the corresponding invariant mass distributions computed
by taking into account this schematic detector resolution.
It is seen that already for a resolution characterized by
$d=15\, MeV$  the $\omega$ peak essentially decreases and practically disappears at
lower resolutions, $d \sim 30\, MeV$.

\section{Summary \label{sumary}}

In summary  we have analyzed the di-electron invariant-mass distribution from Dalitz decays
of ${\eta'}$ mesons produced in $pp$ and $pn$  collisions at intermediate energies.
The corresponding cross section has been
calculated within an effective meson-nucleon approach with
parameters adjusted to describe the free  vector and pseudo-scalar
meson production \cite{ourOmega,ourOmegaFF} in $NN$ reactions near
the threshold. In particular, we found a suitable
parametrization of the cross section for
$pp\to pp\eta'$, which can be used for a prediction of the channel $pn  \to pn\eta'$,
which is important in heavy-ion collisions.
We argue  that by studying
the invariant mass distribution of  the final $e^+e^-$ system from $\eta'$ Dalitz decays
in a large  kinematical interval     one can directly  measure
the  transition form factor $F_{\eta'\gamma\gamma^*}$,  e.g., in $pp$
collisions. Such experiments are
possible, for instance at HADES, CLAS, KEK-PS,
and our results may serve as predictions for these
forthcoming experiments. Experimental information
on form factors is useful for testing predictions of hadronic quantities
in the non-perturbative QCD domain. In particular, as mentioned in Introduction,
the use of the quark flavor basis
in a combined analysis of $\eta$ and $\eta'$ mesons with one mixing angle,
but with topological effects
connected with the $U(1)$ axial anomaly kept, provides
an analytical form of the transition form factors
with two terms, one related to non-strange quarks
another one being entirely expressed via contributions from
strange components \cite{feldmann_FF}. This is in full analogy with the vector meson
dominance with the $\phi$ meson incorporated.
Also, a precise determination of the
nucleon-nucleon-$\eta'$ coupling constant $g_{NN\eta'}$
together with an investigation
of the polarized deep inelastic
lepton scattering, in connection with the Goldberger-Treiman
relation \cite{etaprime_OZI},
can supply additional information on
the flavor-singlet axial constant $G_A(0)$
and on the gluon coupling $g_{NNg}$, hence 
understanding of the origin of the "spin crisis".

We predict the transition form factor and the
invariant mass distribution to
exhibit maxima at di-electron masses close to the $\rho$ and $\omega$ poles,
in contrast to a pure QED calculation
without strong interaction effects which provides a smooth behavior of the
distribution in the whole kinematical range.

\section*{Acknowledgements}
We thank  H.W. Barz and A.I. Titov for useful discussions.
L.P.K. would like
to thank for the warm hospitality in the Research Center Dresden-Rossendorf.
This work has been supported by
BMBF grant 06DR136 and the Heisenberg-Landau program.

 
\end{document}